\documentclass[12pt,twoside,a4paper]{article}
\usepackage{epsfig}
\usepackage{setspace}

\def\beq{\begin{equation}}
\def\eeq#1{\label{#1}\end{equation}}
\def\eeqn{\end{equation}}

\def\beqa{\begin{eqnarray}}
\def\eeqa#1{\label{#1}\end{eqnarray}}
\def\eeqan{\end{eqnarray}}

\let\bar=\overbar

\newcommand{\captionfonts}{\footnotesize}

\makeatletter  
\long\def\@makecaption#1#2{%
  \vskip\abovecaptionskip
  \sbox\@tempboxa{{\captionfonts #1: #2}}%
  \ifdim \wd\@tempboxa >\hsize
    {\captionfonts #1: #2\par}
  \else
    \hbox to\hsize{\hfil\box\@tempboxa\hfil}%
  \fi
  \vskip\belowcaptionskip}
\makeatother   

\def\Title#1{

\setbox0\vbox{\vskip-70pt
{\scriptsize{
{\em
\begin{raggedright}
Primer Encuentro de la Radioastronom{\'{\i}}a Espa{\~n}ola\\
~~~~~~~~~~~~~~~``Memorial Lucas Lara''\\ 
J.C.~Guirado, I.~Mart{\'{\i}}-Vidal, and J.M.~Marcaide (eds.)\\ 
May 9th-11th 2006, Valencia, Spain\\
\end{raggedright}
}}}
}

\wd0=0pt
\ht0=0pt
\box0

\begin{doublespace}
\vspace*{-25pt}
\begin{center} {\Large {\bf #1} } \end{center}
\end{doublespace}

\medskip

}

\newcommand{\runninghead}[2]
{\markboth{\small{\it {#1}}}{\small{\it{#2}}}}

\def\Dslash{\not{\hbox{\kern-4pt $D$}}}
\def\dslash{\not{\hbox{\kern-2pt $\del$}}}

\def\msb{{\bar{\ssstyle M \kern -1pt S}}}

\def\Dm2{\Delta m^{2}}
\def\eV2{\mbox{eV}^{2}}
\def\B8{^{8}\mbox{B}}
\def\Be7{^{7}\mbox{Be}}

\begin{document}


\Title{HIGHEST-RESOLUTION RADIO ASTRONOMY: THE QUEST FOR THE BLACK HOLE}

\begin{raggedright}



J.A.~Zensus, T.P. Krichbaum, S. Britzen\\

{\small{\it Max-Planck-Institut f\"ur Radioastronomie, Auf dem H\"ugel 69, 53 121 Bonn, GERMANY}} \\


\medskip





\runninghead {J. A. Zensus, T.P. Krichbaum, \& S. Britzen}
{Highest-Resolution Radio Astronomy}

\thispagestyle{empty}
\medskip\smallskip


{\small 
{\bf Abstract:} Radio observations with Very Long Baseline Interferometry (VLBI)
provide the highest resolution in astronomy. Combining
earth-bound with space-based telescopes and advancing  the observations 
to mm-wavelengths increases the resolution even further. These methods 
enable us to probe directly the vicinities of the presumed central black 
holes in active galactic nuclei (AGN) and the powerful jets emanating from 
these objects. We provide a brief review of recent results in 
this exciting research domain and we discuss the opportunities for future
work possible with the advent of new instrumental developments.
}

\end{raggedright}
\begin{singlespace}

\section{Studies of Active Galactic Nuclei with VLBI}
In recent years, our view on AGN has changed from being radio-dominated to multi-wavelength. 
With the performance of extensive surveys in different wavebands (e.g. SDSS, deep X-ray surveys), 
the starting of the operation of new earth- and ground-based telescopes (e.g. VLTI, CHANDRA, HESS), 
and the advent of fascinating powerful new observational possibilities like SKA, ALMA, and GLAST in 
the forseeable future, extragalactic astrophysics is faced with unprecedented possibilities for 
research. Different aspects of astrophysics -- formerly strictly separated -- now fruitfully combine 
and provide us with new insights into evolutionary processes in our universe. Active galacti nuclei, 
the growth of the black holes and their evolution within the surrounding galaxies might closely be 
connected with the evolution of the universe itself. An association between galaxy interactions and 
AGN (e.g. Bahcall et al. \cite{bahcall}) seems to be likely although the details of a possible 
AGN-activity cycle are not yet proven. However, it seems possible that e.g. quasar hosts involved 
in recent collisions are not only common, but perhaps ubiquitous. It has been speculated that 
super-massive black holes (SMBH) may be fed or even formed in galaxy-merging events (e.g. 
Kauffmann \& Haehnelt \cite{kauffmann}). \\
Investigating the {\it AGN-phase} requires combining deep interferometric monitoring of the 
radio jets with continuous observations of the flux-density evolution across the wavelength regime 
in order to locate the emission regions.\\ 

\subsection{Pc-scale jets}
In recent years observers have made great progress in mapping (with interferometric techniques)
the morphologies and the evolution of jets in AGN from cm- down to mm-wavelengths and measuring 
the emission over a wide range of luminosities which presently cover eight orders of magnitude. 
Dedicated campaigns for prototypical objects (e.g. Mkn\,421, 3C\,279, S5\,0716+71) have combined 
simultaneous observations across the wavelength regimes with detailed VLBI monitoring of the 
structural evolution. Approaches to also obtain statistically valid kinematic fundamentals for 
AGN led to major surveys, such as the 2cm survey (e.g. Kellermann et al. \cite{kellermann}), 
and the CJF survey at 6cm wavelength (Britzen et al. \cite{cjf}). A comparison of the results 
of the two surveys suggests higher average apparent velocities in the jets of AGN with higher 
observing frequencies (Jorstad et al. \cite{jorstad}, Kellermann et al. \cite{kellermann}, 
Britzen et al. \cite{cjf}). In addition to the observed faster apparent motions, bending seems 
to be more pronounced at higher frequencies (e.g. Britzen et al. \cite{nullvier}). Despite the 
wealth of kinematic information derived so far, it is still unclear how the kinematic properties 
at different radio frequencies are related. \\
In a growing number of jets upstream motions, slow moving and quasi-stationary components trailing 
superluminal features have been observed (e.g. G$\acute{o}$mez et al. \cite{gomez}; Kellermann et al. 
\cite{kellermann}; Britzen et al. \cite {cjf}). Relativistic hydrodynamic simulations of jets 
explain these as a complex combination of phase motions and interactions between perturbations 
and the underlying jet and/or ambient medium. Jet components should no longer be regarded as related 
to fluid bulk motions alone, and {\it component} motions depend on the angular resolution with which 
the jet is observed, since different beam sizes probe different jet structures. The proper 
identification of phase motions requires high time sampling of the jet emission structure. Sparse 
time sampling fails to detect e.g. reconfinement shocks as in 1803+784 (Britzen et al. \cite{mnras}).
Curved jet structures and ``wiggling'' jets seem to be the rule and straight jets the minority 
(e.g. Britzen et al. \cite{cjf}). Some of the bent jets seem to result from helical structure and 
components are ejected at different position angles with time. A precessing jet nozzle is capable 
of explaining these observations and can result from different physical mechanisms, such as a 
precession of the accretion disk (e.g., Linfield \cite{linfield}) or fluid-dynamical instabilities 
in the interface between the jet material and the surrounding medium (Hardee \cite{hardee} and ref. 
therein). Disk precession can be driven by a companion super-massive black hole or another massive 
object (Stirling et al. \cite{stirling}). Alternative possibilities include the precession of the 
disc as an intrinsic property of the accretion system (Lai \cite{lai}; Liu \& Melia \cite{liu}). 
Models of super-massive binary black holes provide an attractive scenario to explain the observed 
properties based on our current concept of hierarchical structure formation in the universe as shown below.     

\subsection{Curved jet structures and Binary Black Holes}
The expected frequent mergers of galaxies over the course of their formation and cosmological evolution 
must lead to the formation of super-massive binary black holes (e.g. Milosavljevic \& Merritt \cite{milos}).
The observation of a luminous accreting binary AGN in the merging galaxy NGC 6240 (Komossa et al. 
\cite{komossa}) can provide an important test of this scenario. Black hole binary mergers could be 
responsible for many or most quasars according to Gould \& Rix \cite{rix}. Detecting more such binary 
systems is therefore of great interest for key topics in astrophysics ranging from galaxy formation to 
activity in galaxies. Here, the expected number of systems based on hierarchical galaxy merger models 
and activity models is large (up to 40\% of bright ellipticals, Haehnelt \& Kauffmann \cite{haehnelt}).
In accordance with other independent groups we presented evidence that these systems can be identified 
by periodicities observable in different wavelength regimes, e.g., repeated flares in the light curves 
(e.g. optical, X-rays, $\gamma$-rays), and helicities in the motions in radio jets as derived from the 
curvature of jet motions (e.g. Britzen et al. \cite{jacques}, Lobanov \& Roland \cite{lobanov}).

\subsection{The unknown}
Despite much observational effort and success in cm- and mm-VLBI, impressive progress in theoretical 
modeling and simulations of magnetohydrodynamical jets, several questions concerning the black 
hole/jet connection remain unsolved:\\
- how are the powerful radio jets of AGN launched?\\
- where is the foodpoint of the jet and how can we detect it?\\
- what is the relation between accretion disk, black hole, magnetic fields and winds?\\
- what is the spin of a black hole (Sgr A* in particular)?\\
- does the spin of the black hole influence the radio jet?\\

\noindent
With future VLBI experiments we can hope to address at least some of these questions.

\section{mm-VLBI: current status and challenges}
Very Long Baseline Interferometry at millimeter wavelengths (mm-VLBI)
allows to image compact galactic and extragalactic radio sources
with micro-arcsecond resolution, unreached by other astronomical observing techniques.
Future global VLBI at short millimeter wavelengths therefore should allow to map
the direct vicinity of the super-massive black holes located at the centers
of nearby galaxies with a spatial resolution of only a few to a few ten gravitational radii.
With the lower intrinsic self-absorption at these short wavelengths, mm-VLBI opens 
a direct view onto the often jet-producing "central engine".\\
In the following we report on new developments in mm-VLBI, with emphasis on 
experiments performed at the highest frequencies possible to date.
We demonstrate that global VLBI at 150 and 230 GHz now is technically feasible
and yields source detections with an angular resolution as high as $25 - 30 \mu$as.
The combination of the existing mm-/sub-mm telescopes 
with future telescopes (e.g. APEX, SMA, CARMA, ALMA, LMT, etc.) will
improve present day imaging capabilities by a large factor. Within the next
decade, one therefore could expect direct images of galactic and extragalactic 
(super-massive) black holes and their emanating outflows.\\

\subsection{Imaging the Jet Base of M87 with 20 ${\bf R_{\rm S}}$}
\begin{figure}[t]
\includegraphics[bb=50 185 470 625,clip=,angle=-90,width=.4\textwidth]{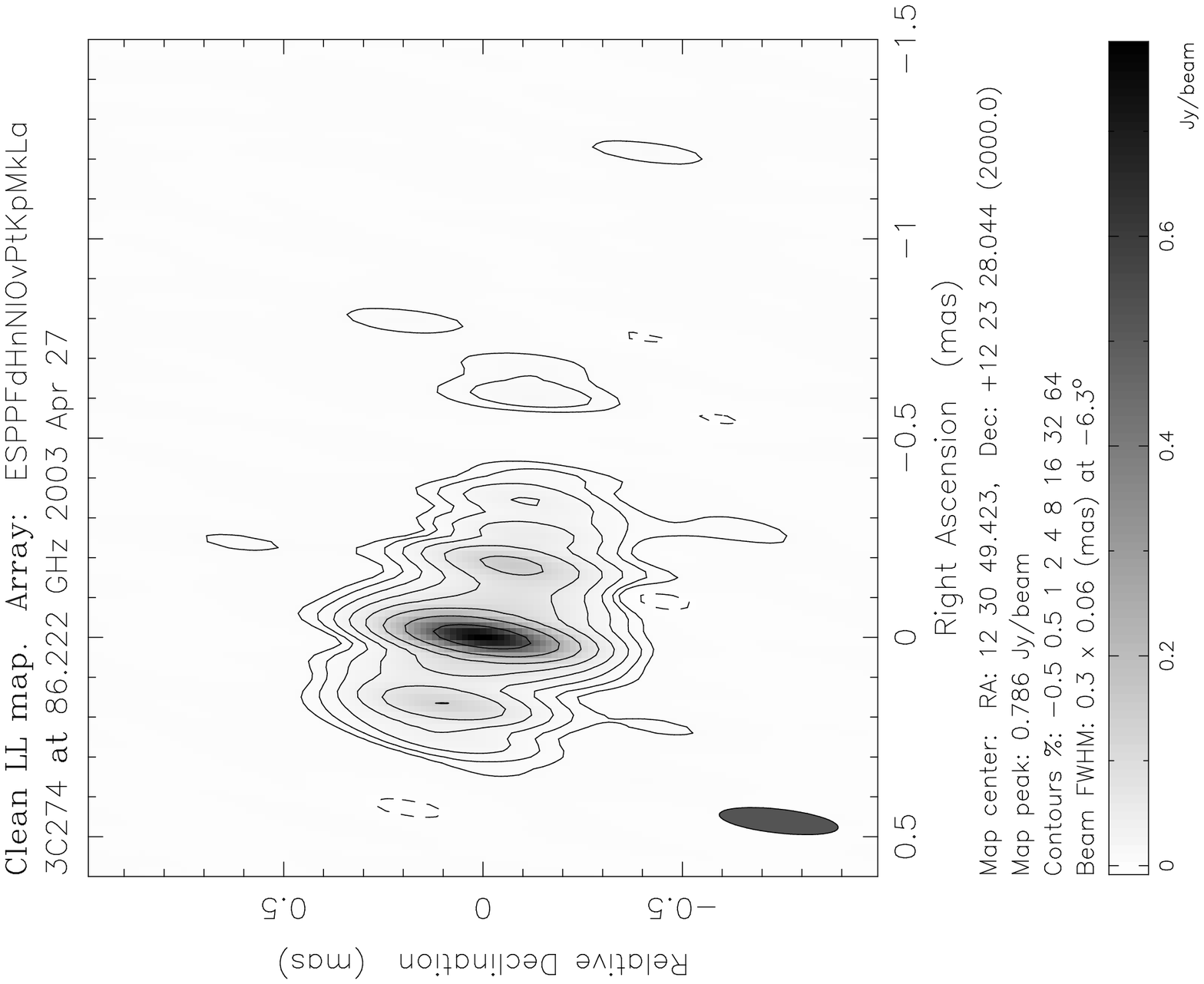}
\hspace{1cm}\begin{minipage}[t]{6cm}
\vspace{0.3cm}
\caption{\label{m87} VLBI image of M87 (3C\,274) obtained in April 2003 at 86 GHz with the Global mm-VLBI 
Array (Krichbaum et al. \cite{krich5}).  Contour levels are -0.5, 0.5, 1, 2, 4, 8, 16, 32, and 64 \% of 
the peak flux of 0.79 Jy/beam. The beam size is 0.30 x 0.06 mas, pa=-6.3$^\circ$. The identification of the 
easternmost jet component as VLBI core or as part of a counter-jet is still uncertain.}
\end{minipage}
\end{figure}

The Global mm-VLBI Array (GMVA)\footnote{web link: http://www.mpifr-bonn.mpg.de/old\_mpifr/div/vlbi/globalmm}
is operational since early 2000. At 86\,GHz ($\lambda=3.5$\,mm), it combines the European antennas 
(Effelsberg 100\,m, Pico Veleta 30\,m, the phased Plateau de Bure Interferometer 6x15\,m, Onsala 
20\,m, Mets\"ahovi 14\,m) with the VLBA. With the participation of the two sensitive IRAM telescopes 
and the 100\,m Effelsberg telescope, the array sensitivity is improved by a factor of $3-4$, when 
compared to the VLBA alone. For compact galactic and extragalactic radio sources, the GMVA provides 
VLBI images with an angular resolution of up to $40$\,$\mu$as. As an example, we show in Fig. 
\ref{m87} a new global-VLBI image of the inner jet of M87 at 86\,GHz (taken from Krichbaum et al. 
\cite{krich5}).  At a distance of 18.7 Mpc, the angular resolution of $300 \times 60$\,$\mu$as 
corresponds to a spatial scale of 30 $\times$ 6 light days, or 100 $\times$ 20 Schwarzschild-radii 
(assuming 3 $\times$ $10^6$ $\rm{M}_\odot$ for the SMBH). The existence of a fully developed jet 
on such small spatial scales gives important new constraints for the theory of jet formation 
and may even indicate rotation of the central SMBH (via comparison with the width of the light 
cylinder).\\

\subsection{VLBI observations of Sgr\,A* at 3, 2, and 1\,mm wavelength}
With VLBI at centimeter wavelengths, the compact radio source Sgr\,A* in the center of our Galaxy 
exhibits a scatter-broadened point like image. Above $\lambda \geq 1.3$\,cm, Sgr\,A* appears 
stationary, i.e. non-variable with time (e.g. Alberdi et al. \cite{alberdi}; Lo et al. \cite{lo93}; 
Lo et al. \cite{lo98}; Marcaide et al. \cite{marcaide}, and references therein). Early VLBI observations 
at  $\lambda =7$\,mm gave first hints that towards shorter wavelengths the source structure may be 
resolved and that the measured source size is larger than the scattering size (Krichbaum et al. 
\cite{krich92}). Better 7\,mm VLBI experiments, performed with more stations and higher sensitivity, 
now confirm an elliptical, point-like brightness distribution, with the major axis of the ellipsoid 
being slightly larger than the scattering size (Lo et al. \cite{lo98}, Bower et al. \cite{bower}). 
Since interstellar scattering effects vanish towards shorter wavelengths, VLBI observations at short 
millimeter wavelength provide the unique opportunity to image the `underlying' and from scintillation 
not affected, {\it intrinsic} source structure.\\

\begin{figure}[t]
\begin{minipage}{18pc}
\includegraphics[width=18pc,angle=-90]{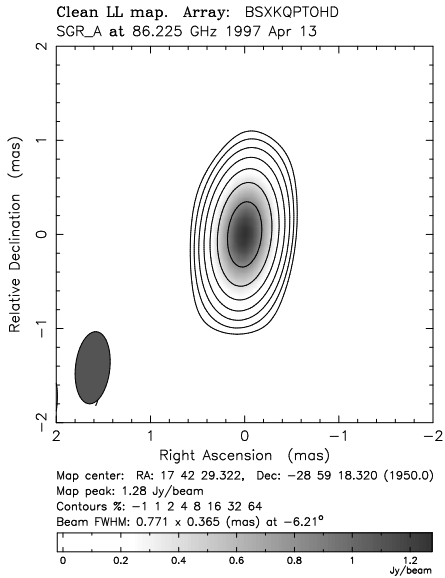}
\end{minipage}
\begin{minipage}[h]{14pc}
\caption{\label{sgra97}Probably the first 86\,GHz VLBI image of Sgr\,A*, as observed in April 1997 (image 
from Krichbaum et al.~\cite{krich6}). The participating antennas are: Effelsberg (100\,m), Pico Veleta (30\,m), 
Haystack (37\,m), Quabbin (14\,m), Pie Town (25\,m), Kitt Peak (12\,m), Owens Valley (5 x 10\,m). A circular 
Gaussian parameterizes Sgr\,A* with flux density $S=1.73 \pm 0.25$\,Jy and size $\theta=0.28\pm0.08$\,mas.}
\end{minipage}
\end{figure}

\begin{figure}[h!]
\begin{center}
\includegraphics[width=23pc,angle=-90,bb=64 39 567 705,clip=]{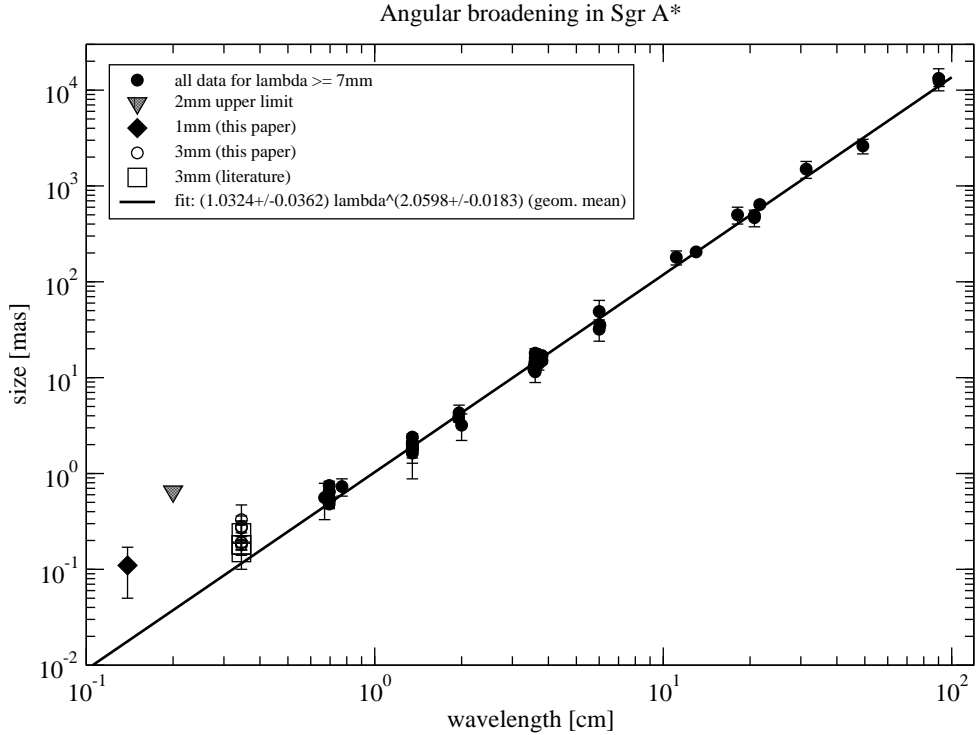}
\caption{\label{size}Angular size of Sgr\,A* plotted versus wavelength. The line denotes a fit 
($a \lambda^\beta$), excluding the short wavelengths $\lambda \leq 7$\,mm, which are not affected by 
interstellar scattering. The fit parameters $a$, $\beta$ are given inside the figure. Published 3\,mm 
measurements are plotted as open squares, new 3\,mm data (Krichbaum et al. \cite{krich6}) by open 
circles. New measurements at 2\,mm (shaded triangle, an upper limit) and 1.4\,mm (diamond) are also shown.}
\vspace*{-1.3cm}
\end{center}
\end{figure}

\noindent
{\bf Results at 3\,mm:}
From the early 1990's onwards, Sgr\,A* was repeatedly observed with VLBI at 3.5\,mm (86\,GHz) in 
various array configurations, using European, American and global VLBI arrays (Krichbaum et al. 
\cite{krich94}; Rogers et al. \cite{rogers}; Doeleman et al. \cite{doeleman}; Shen et al. \cite{shen} 
and reference therein). In Fig. \ref{sgra97} we show one of the early maps obtained from global VLBI 
at 86\,GHz, which reveals a partially resolved point source, similar to the recently published VLBA 
image (Shen et al. \cite{shen}). 
At mm-wavelengths, the accurate determination of flux density and size is often affected by calibration 
uncertainties (low source elevation on the northern hemisphere, atmospheric opacity variations). This 
requires special observing and calibration methods. In parallel to the ongoing research with the VLBA 
(see Shen et al. \cite{shen}), we have continued to observe Sgr\,A*, using the sensitive and large 
antennas at MPIfR (Effelsberg 100\,m telescope) and at IRAM (Pico Veleta 30\,m, Plateau de Bure 6 $\times$ 15\,m). 
With baseline lengths in the range of $\sim 750 - 1700$\,km, and a detection sensitivity of $\sim 50$\,mJy 
($7\sigma$, 512 Mbit/sec) per baseline, this array measures the visibility amplitudes and (closure-) 
phases with particular high $\rm{SNR}$ (in October 2005, Sgr\,A* was detected with $\rm{SNR} \leq  96$).
The measured sizes obtained with this European mm-VLBI array agree well with similar results obtained 
at the VLBA and show a point source (zero closure phase) with a FWHM size in the range of $\sim \theta=0.2\pm 0.05$\,mas, 
clearly indicating that at 86\,GHz ($\lambda=3.5$\,mm) the source is larger than the extrapolated scattering 
size (see Fig. \ref{size}, see also Krichbaum et al. \cite{krich6}).  \\

\noindent
{\bf Results at  2\,mm and 1\,mm:}
VLBI observations at wavelengths shorter than 3\,mm are still difficult and are limited 
by the number of available radio telescopes and their sensitivity. In a pilot VLBI experiment 
at 147\,GHz (2\,mm), Sgr\,A* was detected with a signal-to-noise ratio of $\sim 7$ on the 
short (85\,M$\lambda$) baseline between the Arizona telescopes Kitt Peak (KP, 12\,m) and the 
Heinrich-Hertz telescope (HHT, 10\,m) (Krichbaum et al. \cite{krich2}). This detection 
resulted in an upper limit of the size of $\leq 0.7$\,mas, fully consistent with the measured 
sizes at 3\,mm and 1.4\,mm. So far, all attempts to observe Sgr\,A* with global VLBI at 
wavelengths shorter than 2\,mm failed mainly due to weather and technical problems. This 
leaves the 1.4\,mm VLBI observation performed in March 1995 with the two IRAM telescopes 
still as the only one, in which Sgr\,A* was detected in the 1\,mm band (Krichbaum et al. 
\cite{krich98}). Based on a more accurate knowledge of the total flux density at 215\,GHz, 
Krichbaum et al. \cite{krich6} derived a new value of the source size: $\theta = (110 \pm 60) 
\mu$as. This corresponds to a linear size of only $R_{S} = 11 \pm 6 $ Schwarzschild radii.  
The size measurements at 2\,mm and 1.4\,mm are shown in Fig. \ref{size}. It is obvious that 
future 2\,mm and 1\,mm VLBI experiments are necessary, to confirm these
results and provide more accuracy. If Sgr\,A* can be detected with 1\,mm-VLBI on long, 
inter-continental baselines, images of its intrinsic structure will be made with a spectacular 
angular resolution of only a few Schwarzschild radii.

\section{State of the art: VLBI at 2 and 1\,mm}
A convincing demonstration of the feasibility of VLBI at wavelengths shorter than 3\,mm was made
at 2\,mm (147\,GHz) in 2001 and 2002 (Greve et al. \cite{greve}; Krichbaum et al. \cite{krich2}). 
These first 2\,mm-VLBI experiments resulted in detections of about one dozen quasars on the short 
continental and long transatlantic baselines (participating telescopes: Pico Veleta - Spain; 
Mets\"ahovi - Finland; Heinrich-Hertz and Kitt Peak telescope - Arizona, USA). A big success 
was the detection of 3 quasars on the 4.2\,G$\lambda$ long transatlantic baseline between Pico Veleta 
and the Heinrich-Hertz Telescope: NRAO150 (SNR=7), 1633+382 (SNR=23) and 3C\,279 (SNR=75) (Krichbaum et al. \cite{krich2}).
Motivated by this success, the observations were repeated in April 2003, this time at 1.3\,mm (230\,GHz)
(Krichbaum et al. \cite{krich4}). Now also the phased IRAM interferometer on Plateau de Bure (France) participated. 
On the 1150\,km long baseline between Pico Veleta and Plateau de Bure the following sources were 
detected: NRAO\,150, 3C\,120, 0420-014, 0736+017, 0716+714, OJ\,287, 3C\,273, 3C\,279, and BL\,Lac.
On the 6.4\,G$\lambda$ long transatlantic baseline between Europe and Arizona fringes for
the quasar 3C\,454.3 (SNR=7.3) were clearly seen. For the BL\,Lac object 0716+714, however, only a
marginal detection (SNR=6.8) was obtained. These transatlantic detections
mark a new record in angular resolution in Astronomy (size $< 30 \mu$as). They indicate
the existence of ultra compact emission regions in AGN even at the highest frequencies 
(for 3C\,454.3 at z=0.859, the rest frame frequency is 428\,GHz). So far, we find no evidence for
a reduced brightness temperature of the VLBI-cores at mm-wavelengths, however some variability is 
possible.

\noindent
\section{Future prospects: Approaching the Black Hole}
In order to approach the black hole, the angular resolution will have to be increased even further.
The angular resolution of radio interferometric observations can be improved either by increasing the 
separation between the radio telescopes (longest baseline), or by observing at shorter wavelengths. We 
describe both attempts for the future in the following.\\
\begin{figure}[t!]
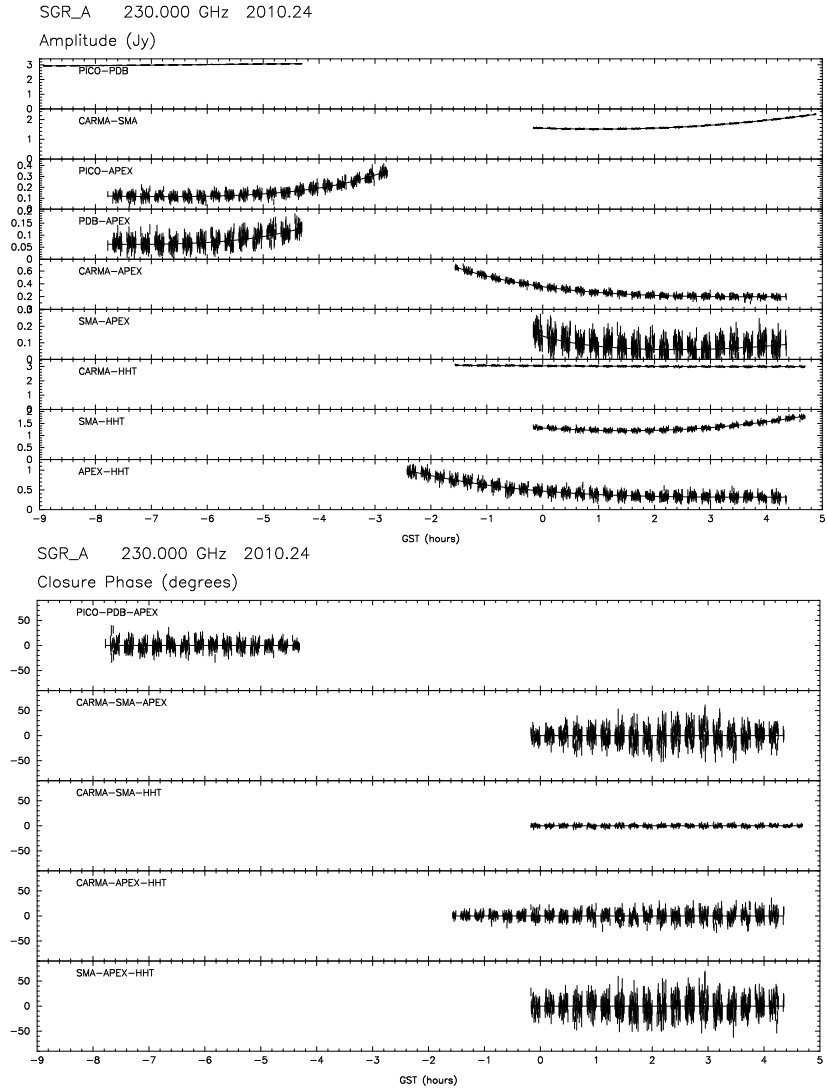

\begin{center}
\hspace*{-4.3cm}
\begin{minipage}{17pc}
\includegraphics[width=17pc,angle=-90]{sgr1mm-vis.epsi}
\includegraphics[width=17pc,angle=-90]{close.epsi}
\end{minipage}
\caption{\label{1mmvis} Simulated visibility amplitudes (top) and closure phase (bottom) for a future 
1.3\,mm VLBI experiment on Sgr\,A* with the following telescopes: Pico Veleta (Spain), Plateau de Bure 
(France), CARMA (California), SMA (Hawaii), HHT (Arizona) and APEX (Chile). For the simulation, a circular 
Gaussian of 2.5\,Jy and $\sim 25$\,$\mu$as FWHM size was assumed.}
\end{center}
\end{figure}

\subsection{mm-VLBI: new telescopes}
Good quality micro-arcsecond resolution VLBI images of the nuclei of galaxies will require an 
increased array sensitivity and better uv-coverage. 
The ongoing development towards observations with much larger bandwidths (several Gbits/s),
and for instantaneous atmospheric phase corrections and coherence prolongation (e.g. via water
vapor radiometry), will further enhance the sensitivity. For sources near the celestial equator and
below (e.g. Sgr\,A*), antennas in the south of Europe and on the southern hemisphere will play an important
role. At 86\,GHz, the addition of Noto, the new 40\,m telescope at Yebes, and the 
Sardinia Radio telescope (D=64\,m) would nicely extend the uv-coverage to the south, largely improving the
imaging capabilities of the GMVA. At higher frequencies ($ > 150$\,GHz), new telescopes like APEX (and later 
ALMA) located in the southern hemisphere and phased interferometers operating as single VLBI antennas 
(PdB, CARMA, SMA) are crucial to provide the necessary resolution and sensitivity for observing 
the 'shadow' of the super-massive BH in the galactic center Sgr\,A*. To illustrate this, we show in Fig. \ref{1mmvis} 
simulated visibility amplitudes and closure phases for a not unrealistic future VLBI observation
of Sgr\,A* at 230\,GHz. With such an experiment one may hope to detect possible deviations of the source
from circular symmetry, as expected for a rotating Kerr black hole.

\subsection{Space-VLBI}
At cm-wavelengths, VLBI with orbiting radio antennas (so called space-VLBI) has revealed high quality
images with an angular resolution 3-4 times higher than achievable by earth-bound VLBI.
One major highlight of the HALCA (VSOP) project certainly is the ability to resolve
prominent quasar jets also in transverse direction (e.g. 3C\,273: Lobanov \& Zensus \cite{zensus}; 
0836+714: Lobanov et al. \cite{lobanov6}), which facilitated a detailed study of jet propagation and 
jet internal processes, like i.e. the development of instabilities. Another important aspect is the 
detection of complex and variable polarization structure, as e.g. seen in 5 GHz VSOP maps of the IDV 
source 0716+714 (Bach et al. \cite{bach}). When extended to shorter wavelengths (higher frequencies) 
space-VLBI experiments, like the planned VSOP2 mission, will provide even higher angular resolution.
At 43 GHz, the angular resolution of VSOP 2 will be of order of a few ten micro-arcseconds,
closely matching the angular resolution from future ground based mm-VLBI at 230 GHz. Images 
obtained at different frequencies and with matched angular resolution, will lead to a determination of the spectral
properties of compact regions, otherwise not observable. Combined with polarimetric imaging
from space and ground, we could hope to obtain definite answers to the question of how jets are
created and accelerated.


\end{singlespace}
\bigskip


\begin{small}

\end{small}
 
\end{document}